\documentclass{article}

\usepackage{PRIMEarxiv}

\usepackage[utf8]{inputenc} 
\usepackage[T1]{fontenc}    
\usepackage{hyperref}       
\usepackage{url}            
\usepackage{booktabs}       
\usepackage{amsfonts}       
\usepackage{nicefrac}       
\usepackage{microtype}      
\usepackage{lipsum}
\usepackage{fancyhdr}       
\usepackage{graphicx}       
\usepackage{amsmath}
\usepackage{bm}
\usepackage{subcaption}
\usepackage{graphicx}

\graphicspath{{media/}}     

\title{Rebound Suppression Mechanisms of Particle-Filled Flexible Shells for Small Body Landings

}

\author{
Tongge Wen\textsuperscript{1,2}, 
Xiaoyu Yang\textsuperscript{2}, 
Sudeshna Roy\textsuperscript{3}, 
Thorsten Pöschel\textsuperscript{3}, 
Xiangyuan Zeng\textsuperscript{2,*} \\
\textsuperscript{1}Department of Engineering Mechanics, Xi'an Jiaotong University, Xi'an 710049, China \\
\textsuperscript{2}School of Automation, Beijing Institute of Technology, Beijing 100081, China \\
\textsuperscript{3}Institute for Multiscale Simulations, Friedrich-Alexander Universität Erlangen-Nürnberg, Cauerstraße 3, 91058 Erlangen, Germany \\
}

\begin{document}
\maketitle

\begin{abstract}
The extremely weak gravity on small bodies ($\sim 10^{-4} \mathrm{~m} / \mathrm{s}^2$) makes landers prone to rebound and uncontrolled drift. To mitigate this, the Hayabusa2 mission employed a particle-filled flexible shell, but the coupled dynamics of shell deformation and internal particle dissipation remain unclear. We develop a computational model representing the flexible shell as a spring-mass network and fully resolve particle collisions, friction, and interactions with granular beds. Results show the flexible shell-granule system dissipates over 90\% of impact energy, far exceeding rigid shells. Energy loss arises from shell–particle coupling, with the particle filling ratio dominating. Impacts on rigid planes produce large shell deformation, while granular beds limit deformation. Scaling and velocity analyses reveal distinct dissipation regimes. These findings clarify energy transfer mechanisms and inform the design of microgravity impact mitigation devices.
\end{abstract}

\section{Introduction}
In small body exploration, landers often adopt hard landing strategies due to extremely low surface gravity ($\approx 10^{-3}-10^{-5} g$) and the absence of dense atmospheres, which precludes stable soft-landing control via thrusters or airbags\cite{Vanwal2018,Vanwal2020,Zeng2022}. Even at low impact speeds, landers can experience multiple rebounds, compromising precise landing and mission objectives. For instance, during the 2014 Rosetta mission, the Philae lander rebounded several times after brief contact with comet 67P and ultimately came to rest in a shadowed cliff crevice, approximately 1 km from the designated site. The originally planned 60-day scientific observations lasted only around 60 hours due to power depletion. This case highlights the inherent limitations of hard landing in microgravity for controlling landing precision. Addressing lander rebound without relying on complex active control remains a critical challenge in small body exploration. In the Japanese Hayabusa2 mission, a new idea was proposed, to adopt a particle damper to reduce rebound\cite{Kusumoto2024}. With this design, the lander achieved only about 15 m deviation from the designated site, showing a significant improvement\cite{Tsuda2020}.

Despite this practical success, the underlying mechanisms of energy dissipation in flexible shell-particle systems remain unclear. Interactions between the deformable shell and enclosed particles form a strongly coupled, nonlinear system, distinct from rigid-shell models. To address this, we develop a coupled flexible shell–particle model using a triangular-mesh spring-mass system, capturing large deformations and nonlinear dynamics efficiently. We investigate the effects of shell stiffness, particle filling ratio, and particle properties on damping performance, and compare impacts on particle layers versus rigid planes. These analyses reveal fundamental mechanisms governing energy dissipation in microgravity particle-damping systems.

\section{Methodology}
\label{sec2}

\begin{figure}[htbp]  
    \centering
    \includegraphics[width=0.6\textwidth]{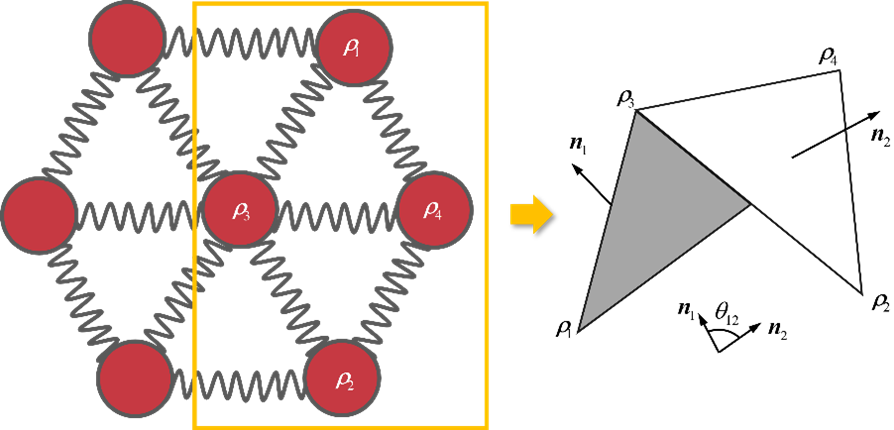} 
    \caption{Discretized mass–spring representation of the flexible shell.}  
    \label{fig1}         
\end{figure}

In this paper, a spring-mass model is employed to represent the flexible shell structure. The continuous shell is discretized into a regular grid of mass points, with each node treated as a particle possessing mass. Adjacent nodes are connected by springs, forming an elastic network, as illustrated in Figure \ref{fig1}. The shell’s tensile deformation is described using Hooke’s law\cite{Gtz2023a,Gtz2023b}:

\begin{equation}
\boldsymbol{F}_i^{\text {spring }}=-k_s\left(\boldsymbol{\rho}_{i j}-\boldsymbol{\rho}_{i j}^0\right) \boldsymbol{\rho}_{i j}+2 \gamma_s \sqrt{k_s m_{\text {eff }}} \dot{\boldsymbol{\rho}}_{i j}
\end{equation}

where ${\boldsymbol{\rho}}_{i j}$ denotes the relative position between nodes $i$ and $j$, ${\boldsymbol{\rho}}_{i j}^0$ is the spring’s rest length, $\dot{\boldsymbol{\rho}}_{i j}$ is the relative velocity, $k_s$and $\gamma_s$ are the stiffness and damping coefficients, respectively, and $m_{\text {eff }}$ is the effective mass.

Bending forces on the shell are represented by changes in the angles between the normals of adjacent triangular elements, decomposed into elastic forces $\boldsymbol{F}_i^{\text {elastic }}$ and dissipative forces $\boldsymbol{F}_i^{\text {dissipative }}$, which approximate the shell’s bending stiffness. Accordingly, the total force on the $i$-th shell node can be expressed as:

\begin{equation}
\left\{\begin{array}{l}
\boldsymbol{F}_i^{\text {elastic }}=-k_e \frac{\left|\boldsymbol{\rho}_4-\boldsymbol{\rho}_3\right|^2}{\left|\boldsymbol{n}_1\right|+\left|\boldsymbol{n}_2\right|}\left(\sin \frac{\theta_{12}}{2}-\sin \frac{\theta_{12}^0}{2}\right) \boldsymbol{u}_i \\
\boldsymbol{F}_i^{\text {dissipative }}=-k_d\left|\boldsymbol{\rho}_4-\boldsymbol{\rho}_3\right| \dot{\theta}_{12} \boldsymbol{u}_i
\end{array}\right.
\end{equation}

where $k_e$ and $k_d$ are material parameters associated with bending, and $u_i$ is a linear combination of the normal directions of adjacent triangles, given by:

\begin{equation}
\left\{\begin{array}{l}
\boldsymbol{u}_1=\left|\boldsymbol{\rho}_4-\boldsymbol{\rho}_3\right| \frac{\boldsymbol{n}_1}{\left|\boldsymbol{n}_1\right|^2} \\
\boldsymbol{u}_2=\left|\boldsymbol{\rho}_4-\boldsymbol{\rho}_3\right| \frac{\boldsymbol{n}_2}{\left|\boldsymbol{n}_2\right|^2} \\
\boldsymbol{u}_3=\frac{\left(\boldsymbol{\rho}_1-\boldsymbol{\rho}_4\right)\left(\boldsymbol{\rho}_4-\boldsymbol{\rho}_3\right)}{\left|\boldsymbol{\rho}_4-\boldsymbol{\rho}_3\right|} \frac{\boldsymbol{n}_1}{\left|\boldsymbol{n}_1\right|^2}+\frac{\left(\boldsymbol{\rho}_2-\boldsymbol{\rho}_4\right)\left(\boldsymbol{\rho}_4-\boldsymbol{\rho}_3\right)}{\left|\boldsymbol{\rho}_4-\boldsymbol{\rho}_3\right|} \frac{\boldsymbol{n}_2}{\left|\boldsymbol{n}_2\right|^2} \\
\boldsymbol{u}_3=\frac{\left(\boldsymbol{\rho}_1-\boldsymbol{\rho}_3\right)\left(\boldsymbol{\rho}_4-\boldsymbol{\rho}_3\right)}{\left|\boldsymbol{\rho}_4-\boldsymbol{\rho}_3\right|} \frac{\boldsymbol{n}_1}{\left|\boldsymbol{n}_1\right|^2}+\frac{\left(\boldsymbol{\rho}_2-\boldsymbol{\rho}_3\right)\left(\boldsymbol{\rho}_4-\boldsymbol{\rho}_3\right)}{\left|\boldsymbol{\rho}_4-\boldsymbol{\rho}_3\right|} \frac{\boldsymbol{n}_2}{\left|\boldsymbol{n}_2\right|^2}
\end{array}\right.
\end{equation}

These forces are then applied to the vertices, and by integrating the vertex motion, we can obtain the overall deformation of the shell.

\section{Results}
\begin{figure}[htbp]
    \centering

    \begin{subfigure}[b]{0.6\textwidth}
        \centering
        \includegraphics[width=\textwidth]{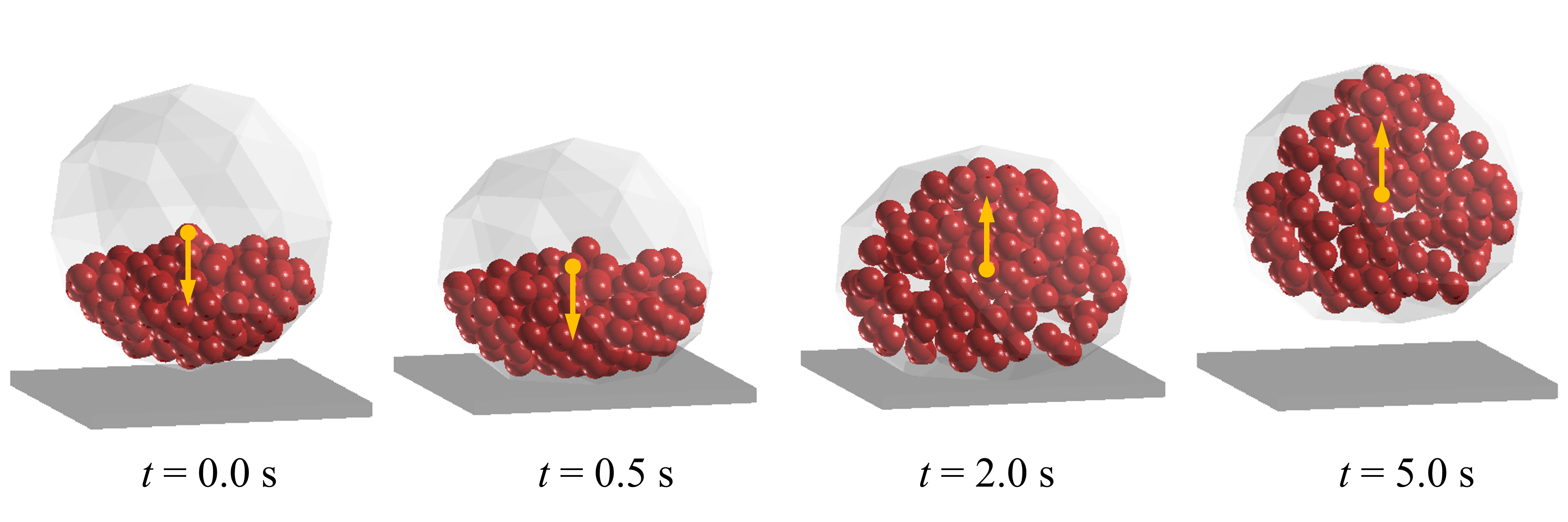}
        \caption{Rigid shell.}
        \label{fig2a}
    \end{subfigure}

    \vspace{0.5em} 

    \begin{subfigure}[b]{0.6\textwidth}
        \centering
        \includegraphics[width=\textwidth]{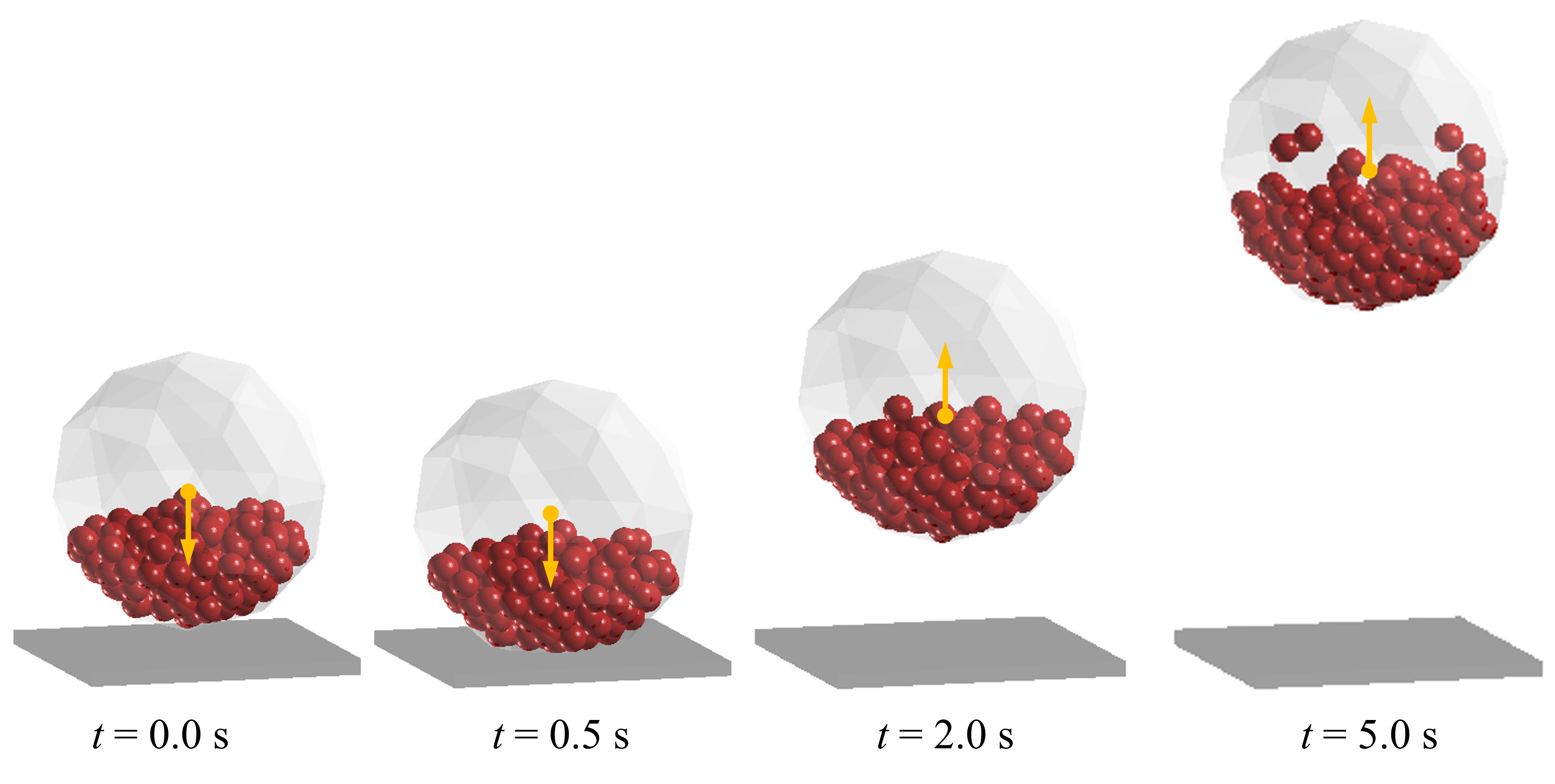}
        \caption{Flexible shell.}
        \label{fig2b}
    \end{subfigure}

    \caption{Particle damper with rigid and flexible shells impacting the rigid surface.}
    \label{fig2}
\end{figure}

We first compare the damping performance of a fully rigid shell and a flexible shell during impact with the ground. Here, the rigid shell is defined as one that does not deform at all upon impact. Figure \ref{fig2} shows the impact process of the shell with a 30\% filling ratio. For both cases, the particle filling ratio is varied from 10\% to 50\%, with 50\% representing nearly full packing.

\begin{figure}[htbp]  
    \centering
    \includegraphics[width=0.6\textwidth]{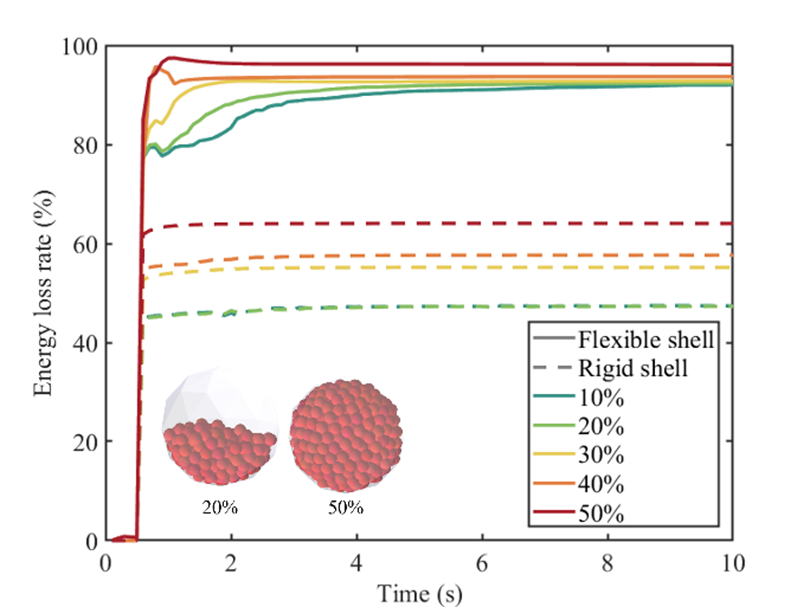} 
    \caption{Impact energy loss rate of particle-filled shells as a function of the particle filling ratio.}  
    \label{fig3}         
\end{figure}

Figure \ref{fig3} illustrates the post-impact energy loss rate for the rigid shell and flexible shell. Interestingly, the rigid and flexible shells exhibit markedly different damping behaviors. For the rigid shell, energy dissipation is dominated by particle collisions. Damping efficiency increases with the filling ratio, reaching a maximum near 50\%, where particle–particle collisions are most frequent and momentum transfer is strongest. In this case, the primary sources of energy loss are inelastic particle collisions and frictional contacts within the damper. By comparison, the flexible shell exhibits a coupled shell–particle damping mechanism. The shell deforms, storing and releasing energy during impact, which interacts with the internal granular motion. Consequently, energy is dissipated not only through particle–particle collisions but also via particle–flexible shell friction and material hysteresis of the shell itself. This coupling yields consistently high energy dissipation, typically exceeding 90\% across different filling ratios. While damping slightly increases with more particles, it is far less sensitive to the filling ratio than in the rigid case. These results indicate that rigid shells rely primarily on collision-driven dissipation, where more particles enhance energy loss, whereas flexible shells rely on deformation-coupled dissipation, with shell–particle interactions dominating and ensuring stable, high damping efficiency.

\begin{figure}[htbp]  
    \centering
    \includegraphics[width=1.0\textwidth]{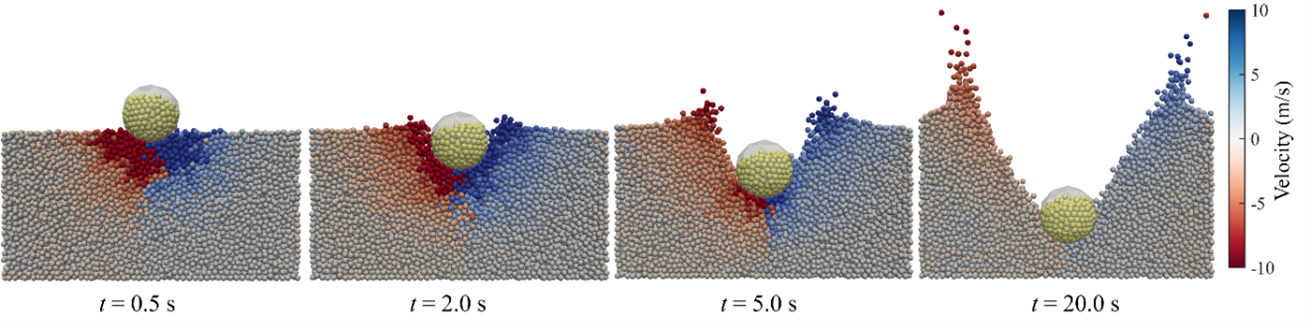} 
    \caption{Flexible shell impact on the granular bed.}  
    \label{fig4}         
\end{figure}

Figure \ref{fig4} shows a particle-filled flexible shell impact on the granular bed. It can be seen that impacts on a rigid plane produce large shell deformation, while impacts on a granular bed limit deformation, highlighting distinct response regimes. Scaling analysis further reveals that penetration depth scales with drop distance with an exponent of $\sim 0.18$, much smaller than the 1/3 law observed for rigid spheres. The velocity decay exhibits a clear two-stage behavior: rapid exponential reduction at the initial impact, followed by a quasi-steady linear decay, indicating a shift in the dominant dissipation mechanism.

\section{Conclusion}
This study demonstrates the distinct damping mechanisms of rigid and flexible particle-filled shells in microgravity impacts. Rigid shells rely primarily on collision-driven dissipation, with energy loss increasing with particle number, while flexible shells exhibit coupled shell–particle damping, where shell deformation interacts with internal granular motion to produce consistently high energy dissipation ($>$90\%) across varying filling ratios. Impacts on granular beds limit shell deformation and produce two-stage velocity decay, highlighting the influence of surface compliance on dissipation dynamics. Overall, flexible shells provide stable and efficient damping, offering a promising passive strategy to mitigate lander rebound in small body exploration.

\section*{Acknowledgments}
This work is supported by the National Natural Science Foundation of China (No. 12222202).

\bibliographystyle{unsrt}  
\bibliography{references}

\end{document}